\documentclass[aps.prl,reprint,groupedaddress]{revtex4-1}

\usepackage{amsmath}
\usepackage{amssymb}
\usepackage{graphicx}
\usepackage{float}
\usepackage{subfloat}

\begin{document}


\title{ Pinning Susceptibility: The effect of dilute, quenched disorder on jamming}


\author{Amy L. Graves {\small(formerly, Bug)}$^{1}$, Samer Nashed$^1$, Elliot Padgett$^{1,2}$, Carl P. Goodrich$^3$, Andrea J. Liu$^3$ and James P. Sethna$^4$}

\affiliation{ $^1$ Dept. of Physics and Astronomy, Swarthmore College, $^2$ School of Applied and Engineering Physics, Cornell University, $^3$ Dept. of Physics and Astronomy, University of Pennsylvania, $^4$ Department of Physics, Cornell University }


\date{\today}

\begin{abstract}
We study the effect of dilute pinning on the jamming transition.  Pinning reduces the average contact number needed to jam unpinned particles and shifts the jamming threshold to lower densities, leading to a pinning susceptibility, $\chi_p$.  Our main results are that this susceptibility obeys scaling form and diverges in the thermodynamic limit as $\chi_p \propto |\phi - \phi_c^\infty|^{-\gamma_p}$ where $\phi_c^\infty$ is the jamming threshold in the absence of pins.  Finite-size scaling arguments yield these values with associated statistical (systematic) errors  $\gamma_p  = 1.018 \pm 0.026 (0.291) $ in $d=2$ and $\gamma_p =1.534 \pm 0.120 (0.822)$ in $d=3$.  Logarithmic corrections raise the exponent in $d=2$ to close to the $d=3$ value, although the systematic errors are very large.
\end{abstract}


\maketitle

In jammed packings of ideal spheres, particles are locked into position by their repulsive interactions with their neighbors, which in turn are locked into position by their neighbors, and so on, so that the entire system is mechanically stable.  Pinning is an alternate way of locking a particle into place, so the interplay of pinning and jamming can potentially lead to interesting new behavior.  Pinning is known to have a rich interplay with glassiness; pinning  raises the glass transition~\cite{Kim2003} and can be used to probe its nature~\cite{Cammarota2012,Berthier2012} and associated length scales~\cite{Berthier2012,Karmakar2012}. In jammed systems, pinning lowers the jamming density~\cite{Reichhardt2012,Brito2013} and allows access to length scales~\cite{Mailman2012}. Here we show that the addition of quenched disorder in the form of random pinning has a singular effect on jamming.  In the dilute pinning limit, jamming is highly susceptible to pinning, with a ``pinning susceptibility" that diverges at the jamming transition as a power law in the thermodynamic limit.
	
In spin systems such as the Ising model, the magnetic susceptibility can be calculated by considering the response to ``ghost spins" or especially-designated spins~\cite{Griffiths1967} in the limit that their density vanishes; similarly, in percolation or correlated percolation one can calculate susceptibilities to ``ghost" sites or bonds~\cite{Reynolds1977,Schwarz2006} that are vanishingly probable.  Here, we consider the response to ``ghost pins." Systems of $N$ particles, of which a fraction $n_f$ are pinned, are prepared by quenching infinitely rapidly from infinite temperature, $T=\infty$, to $T=0$ at a volume fraction $\phi$.  We calculate the fraction of such systems that are jammed, or equivalently, the probability that a state prepared in such a way is jammed, $p_j(\phi, N, n_f)$.  We then define the pinning susceptibility in the limit of vanishing pinning:
\begin{equation}
\chi_p= \lim_{n_f \to 0} \frac{\partial p_j(\phi,N, n_f)}{\partial n_f}
\label{chi_p}
\end{equation} 
We find that $\chi_p(\phi, N)$ and the probability of being jammed, $p_j(\phi, N, n_f)$ obey scaling form and that $\chi_p$ diverges  in the infinite size limit.		

There have been two distinct approaches to studying scaling near the
jamming transition: in terms of a configuration-dependent or
infinite-system critical point. Each finite jammed configuration of particles,
$\Lambda$,  has its own
critical volume fraction $\phi_ c^\Lambda$, which converges to a single value
 $\phi_c^\infty$
only for infinite system sizes~\cite{epitome2003, Chaudhuri2010, Vagberg2011}. 
For many purposes, scaling
behavior near jamming is best done by measuring the deviation 
from the configuration-dependent critical density (as suggested by
Refs.~\cite{OHern2002,epitome2003}). Here, since we are studying the convergence
of the distribution of the configuration-dependent critical densities to the
infinite-system critical density, we naturally make use of the other
approach, scaling in terms of deviation from $\phi_c^\infty$. 
The existence of two distinct scaling pictures 
is seen in many other systems with sharp, global transitions in behavior, as originally discovered in the depinning of charge-density waves~\cite{MyersS1993A,MyersS1993B,Middleton1993}.
Such systems may not obey the inequality between the correlation length and dimension $\nu \ge 2/d$ derived for equilibrium
systems, unless analyzed using deviations from the infinite-system critical
point~\cite{PazmandiSZ97, Chayes1986}.

To study the pinning susceptibility, we conducted numerical simulations on packings of $N$ repulsive soft spheres in $d$ dimensions~\cite{OHern2002,epitome2003} at fixed area (two dimensions, $d=2$) or volume ($d=3$) in a square ($d=2$) or cubic ($d=3$) box with periodic boundary conditions.
We considered 50:50 mixtures of disks ($d=2$) or spheres ($d=3$) with a diameter ratio of 1.0:1.4.   Particles $i$ and $j$ with radii $R_i$ and $R_j$ interact with pairwise repulsions
\begin {equation} U(r_{ij}) = \frac{\varepsilon}{\alpha}\left(1-\frac{r_{ij}}{R_i+R_j}\right)^\alpha \  \Theta\left(1-\frac{r_{ij}}{R_i+R_j}\right)
 \label{eq:pairwise_interaction}	
 \end{equation}
with  $\alpha = 2$ (harmonic)  or $ \frac{5}{2}$  (Hertzian) .  
\footnotetext{formerly Amy L. R. Bug}

The initial position of each particle was generated randomly, and the positions of $N_f=n_f N$ particles (chosen at random) were fixed.  Configurations in which fixed particles overlap were excluded. We then minimized the energy of the system to obtain jammed packings. The upper inset of Fig. 2 contains a sample configuration for $N=256, N_f = 2$.

We calculated the jamming probability  $p_j(\phi, N, n_f)$ at $T=0$ for systems of size $N =  600, 1000, 2000$ and $4000$ in $d=2$,  and $N = 800, 1600, 2400$ and $3200$ in $d=3$. For the small fractions of fixed particles studied here, the criterion for judging a system to be jammed is the same as in previous studies~\cite{Goodrich2012}: jammed systems have positive bulk moduli, energies and pressures. 
 
As for systems with no pinned particles, we find that systems with dilute random pinning are isostatic at the onset of jamming.  In agreement with earlier studies on jammed hard sphere systems with dilute pinning~\cite{Brito2013}, we find that dilute pinning can result in a generalized isostatic condition.  One must distinguish between two types of contacts:  the number $N_{mm}$ between two particles which are mobile during equilibration, and  the number, $N_{mf}$, between one mobile and one fixed particle.   Each of the $N_f$ fixed particles requires no contacts to be stable, while each of the $N_m = N - N_f$ mobile particles require, on average, a minimum of $ Z_{iso}$ contacts. When $N_f = 0$,  $N_m = N$ and $Z_{iso} = 2d - \frac{2d}{N_m} + \frac{2}{N_m}$, where the second term arises from $d$ zero modes associated with global translations allowed by translational invariance, and the third term is needed to ensure a nonzero bulk modulus~\cite{Goodrich2012}.   Our equilibration protocol breaks translational invariance  when $N_f \ge 1$; thus in this case  $Z_{iso} = 2d + \frac{2}{N_m}$.   Since the average number of contacts for a mobile particle is $Z_m = (2 N_{mm} + N_{mf})/N_m$, the number of excess contacts that constrain mobile particles is  $N_{excess} = N_m(Z_m - Z_{iso})$, or 
\begin{equation}  
N_{excess} = N_{mm} + N_{mf} - d N_m + dq - 1
\label{excess}
\end{equation}
where  $q = 1$ for $N_f = 0$, and $q = 0$ for $N_f > 0$. 
Fig. 1a shows that this relationship is upheld: isostaticity means that the excess number of contacts approaches zero as $p \rightarrow 0$.  Additionally, Fig. 1a shows scaling collapse onto universal curves as  function of rescaled pressure, $p^{1/2} N$.  This is exactly the same as what is observed for systems without pins ~\cite{Goodrich2012}.  Note that Fig.~1a is analogous to Fig. 2c of Ref.~\cite{Goodrich2012} in the absence of pins, which shows not only the region of slope $1$, but crossover to a slope of $2$ at very low pressures, arising from a Taylor expansion  of $(Z - Z_{iso})$ in $p$ for finite systems. In Fig.~1a there is perhaps the hint of a crossover to a higher slope at $p^{1/2}N \lesssim 1$, but the data are quite noisy at such low pressures.

\begin{figure}[htpb]
 \centering
 \includegraphics[width=3.4in]{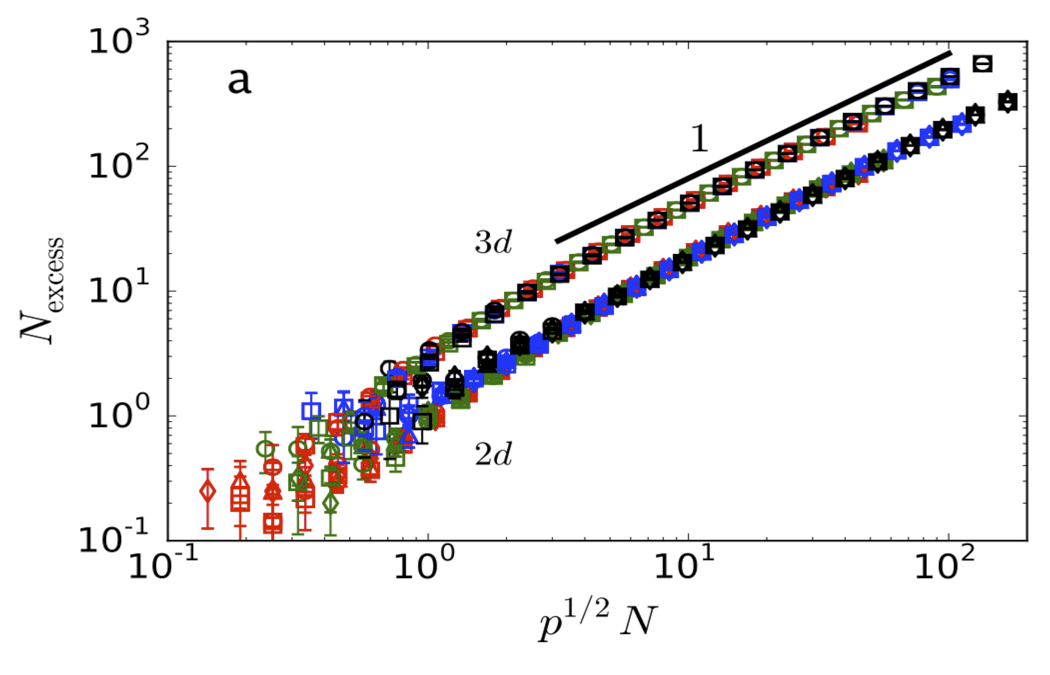}
 \includegraphics[width=3.5 in]{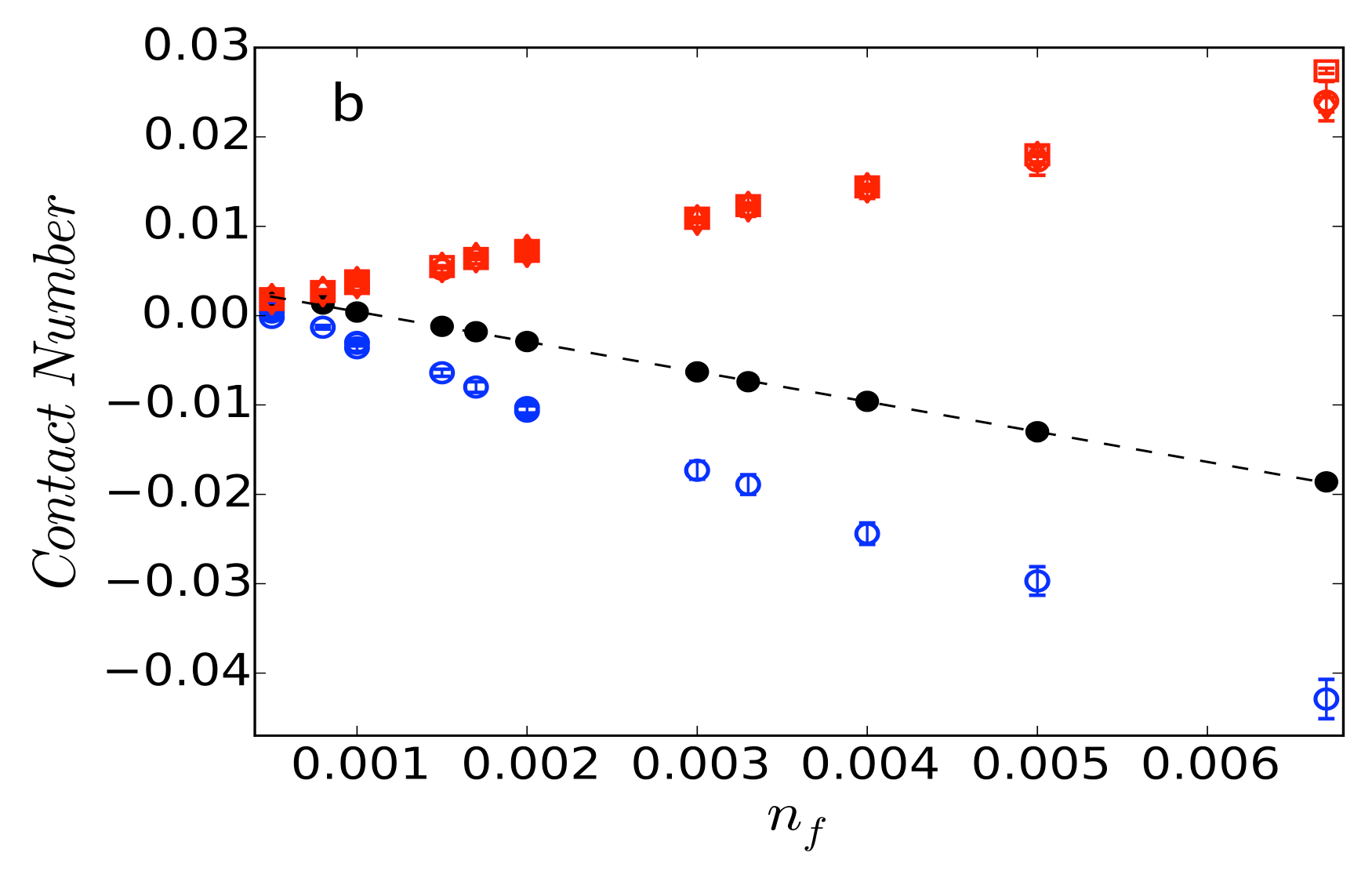}
\label{Fig_excess_various}
 \caption{a: Excess contacts as a function of rescaled pressure.   $d=2$ and  $d=3$ systems are as marked. $N=600, 1000, 2000,  4000$  colors are red, green, blue, black.   $N= 800, 1600, 2400, 3200$  colors are red, green, blue, black.  Symbols for $N_f = 1, 2, 3, 4$
are square, circle, triangle, and diamond.   b: Contributions to contact number for mobile particles.  $Z_{mm} - 4.0$ are shown as blue, $Z_{mf}$  as red.  Symbols for $log_{10} \ p = -6,-5, -4 $ are circles, triangles, squares.  Filled black circles fit by dashed line are $Z_{m } - 4.0$
 for $log_{10} \ p = -6 $ .}
\end{figure}

One might expect that since pinned particles support mobile ones, the number of mobile contacts will decrease with increasing pin density. Indeed, for all pressures studied,  increasing $n_f$ decreases the average number of mobile contacts, $Z_m$. (The only exception is a slight, but nevertheless reproducible, uptick between $N_f =0, 1$, related to the loss of translational invariance.)
Fig. 1b shows $Z_m$ broken down into $Z_{mm}$ arising from mobile-mobile contacts, $Z_{mm} \equiv 2 N_{mm} / N_m$, and $Z_{mf}$ from mobile-fixed contacts. At a pressure low enough to approximate the jamming threshold (circles), the average contact number $Z_m(n_f)$ (filled circles)  is well-fit by the dashed line shown.  Interestingly, raising the pressure by a couple of orders of magnitude does not result in significant changes in $Z_{mf}(n_f)$ (red symbols).    In Fig. 1b, $Z_{mm} - 4.0$ is contrasted with $Z_{mf}$ to show that mobile-mobile contacts disappear more rapidly than mobile-fixed contacts replace them.  Thus, jamming in the presence of pinned particles is an unexpectedly ``frugal'' process, in terms of its use of mobile particles to produce global mechanical stability.

Increasing $n_f$ raises the probability of jamming at any given value of $\phi$,  in accord with previous work on jamming in the presence of fixed particles \cite{Reichhardt2012, Brito2013}.  Increasing $N$ steepens the jamming probability, as in the absence of pinning~\cite{epitome2003}.   These features are illustrated in Fig. 2, which shows the  jamming probability  $p_j(\phi, N, n_f)$ averaged over 10,000-30,000 $d=2$ systems of size $N =  600$ and $2000$, with harmonic repulsions and $N_f = 1,2, 3$ and $4$ fixed particles. The dashed lines in Fig. 2 are fits to a two-parameter logistic sigmoidal form:
\begin{equation}
p_j(\phi, N, n_f) \equiv \frac{1}{(1+e^{a(-\phi+b)})}
\label{pj}
\end{equation}  
where $a(N,n_f)$ is the ``width'' of $p_j$, in that it spans probabilities from $\frac{1}{4}$ to $\frac{3}{4}$; while $b(N,n_f)$ is the value of volume fraction at which $p_j = \frac{1}{2}$.   For all $N$ and $n_f$ studied, logistic sigmoid fits to  $13-21$ independent $\phi$ values result in $\chi^2$ values less than 0.10. A slightly more flexible three-parameter fit, to ``Richard's curve",  does not yield significantly better measures of goodness-of-fit.   

\begin{figure}[h]
\centering
\includegraphics[width=3.4in]{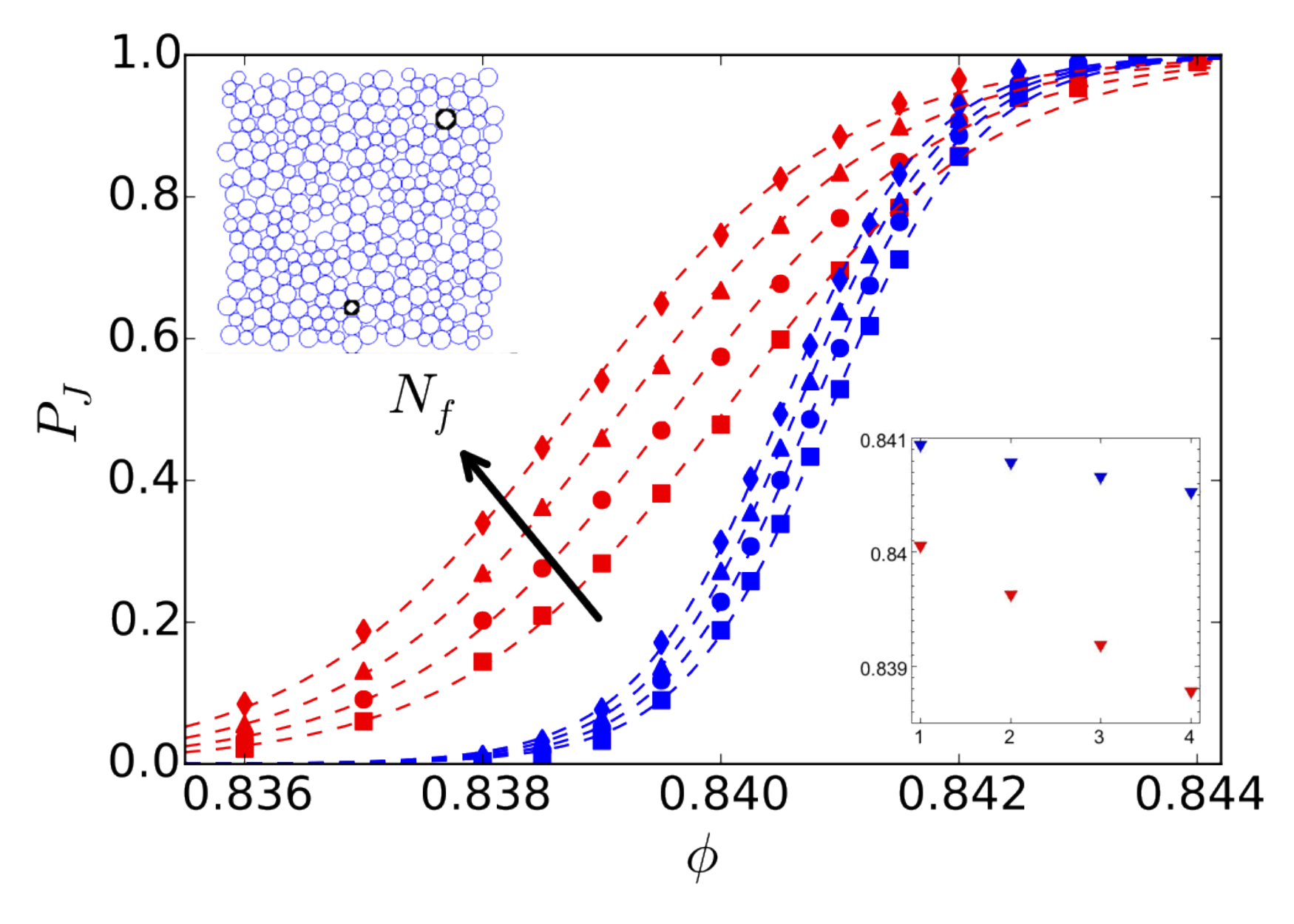}
\caption {Probability $p_j$ of  jamming as a function of packing fraction, $\phi$, in $d=2$ systems for $N=600$ (red) and $N=2000$ (blue).  $N_f = 1, 2, 3, 4$ are represented by square, circle, triangle, and diamond symbols, respectively.  Dashed curves through data are fits to a logistic sigmoid function. Upper inset: An equilibrated configuration in $d=2$ with $N_f = 2$.  Lower inset: Value of $\phi$ such that $p_j = 1/2$  for $N=600$ (red) and $2000$ (blue) versus $N_f$,  the number of pinned particles.}
\label{p_j}
\end{figure}

We now propose a scaling ansatz for $p_j$.    Since  the fraction of pinned sites, $n_f $, is an independent control parameter with which to approach $\phi_c^\infty$,  a two-variable scaling function can be constructed for the jamming probability.  There is significant evidence that the upper critical dimension of the jamming transition is $d_c=2$~\cite{Goodrich2012,Charbonneau2012, Charbonneau2014,Charbonneau2015}.  For $d \ge 2$, we therefore expect that finite scaling depends not on linear system size, $L$, but on particle number $N$~\cite{Binder1985}.  We therefore propose

\begin{equation}
p_j = F( \Delta \phi \ N^{\upsilon} , n_f  N^{\gamma_p \upsilon} ) 
\label{two_variable}
\end{equation}
where $\Delta \phi$ is the distance from the jamming transition for the unpinned, infinite system: $ \Delta \phi= \phi - \phi_c^\infty$.

We can rewrite Eq.~\ref{pj} in terms of the scaling variable $x \equiv \Delta \phi \ N^{\upsilon}$ as 
\begin{equation}
p_j =  \frac{1}{(1+e^{-\tilde{a} x N^{-\upsilon} - \tilde{b}}) }
\label{p_jScaled}
\end{equation}
with 
$ \tilde{a} = a $ and $\tilde{b} = \phi_c^\infty - ab $ .
From the logistic sigmoid fits we can obtain functions $\tilde{a}(y), \ \tilde{b}(y)$, critical exponents $\gamma_p$, $ \upsilon$, and the jamming threshold  $\phi_c^\infty$, where $y \equiv n_f  N^{\gamma_p \upsilon}$. Because the pinning susceptibility is defined in the dilute pinning limit, we are interested in the limit $n_f \rightarrow 0$, or $y \rightarrow 0$.
(We note that in fitting these quantities to our numerical data on $p_j$ vs. $\phi$, the limit $n_f \rightarrow 0$ is taken as the limit $N_f \rightarrow 1$ and not $N_f \rightarrow 0$ since one pin destroys translational invariance. The distinction between $N_f=0$ and $N_f=1$ is irrelevant in the limit $N \rightarrow \infty$.)  We seek the behavior of $\tilde{a}(y)$ and $\tilde{b}(y)$ near $y=0$:
\begin{equation}
\tilde{a} = a_0 + a_1 y  \ ; \ \ \tilde{b} =  b_0 + b_1 y
\label{aAndb}
\end{equation}
with higher-order terms in $y$ neglected.  Table I shows the parameters  in $p_j$ from nonlinear least square fitting in $d=2$ (four system sizes, four pinning densities) and $d=3$ (four system sizes, two pinning densities). 

\begin{table}[htdp]
\label{paramTable}
\begin{center} \begin{tabular}{|c||c|c|c|c|c|c|} \hline
  & $d=2$ value & error$^a$ & error$^b$&  $d=3$ value & error$^a$ & error$^b$\\ \hline 
$\upsilon$ & 0.491 & 0.004 & 0.045 & 0.439 & 0.007 & 0.051  \\ \hline
 $\gamma_p$ & 1.018 & 0.026 & 0.291& 1.534 & 0.120 & 0.822\\  \hline 
 $\phi_c^\infty$ &0.8419 &$<$0.0001 & 0.0001 & 0.6472 & 0.0001 &0.0004 \\  \hline 
 $a_0$ & 38.555 & 1.088 & 12.329 & 46.177 & 2.558 & 17.521 \\ \hline 
 $b_0$ & 1.648 & 0.017 & 0.193 & 3.370 & 0.065 & 0.442 \\ \hline 
 $b_1$ & 9.646 & 0.879 & 9.958 & 2.970 & 1.145 & 7.844\\ \hline
\end{tabular} \caption{Best fit parameters for Eqs.~\ref{p_jScaled},~\ref{aAndb}.
 $^a$Traditional statistical errors.
 $^b$Bounds roughly incorporating systematic errors. Estimated from deviations between model and data \cite{Mortensen2005}, systematic variance is approximated
as twice the best-fit $\chi^2$ divided by the number of parameters.  
}
\end{center}
\end{table}
The parameter $a_1$ is sufficiently close to zero that it is not listed in Table I.  The widths of the sigmoids do not vary significantly with $N_f$; the principal result of increasing $N_f$ is move the sigmoid to the left, leading to jamming at lower values of $\phi$ (as in lower inset of Fig. 2).

The scaling ansatz for the jamming probability, Eq.~\ref{two_variable}, suggests a one-variable scaling ansatz for the pinning susceptibility,
\begin{equation}
\chi_p (\Delta\phi, N) = | \Delta\phi|^{-\gamma_p} \  g_d(\Delta \phi N^{\upsilon})  .
\label{scaling_function}
\end{equation}
Note that we have an explicit form for the function $g_d$, where $d \geq 2$ is the dimensionality,
via Eqs. 1, \ref{p_jScaled} and \ref{aAndb}:
\begin{equation}
g_d(x) = \frac{b_1 e^{b_0 + a_0 \  x}}{(1 + e^{bo + ao \  x})^2}
\label{g} 
\end{equation}

\begin{figure}[htpb]
\centering
\hspace{0.1in}\includegraphics[width=3.2in]{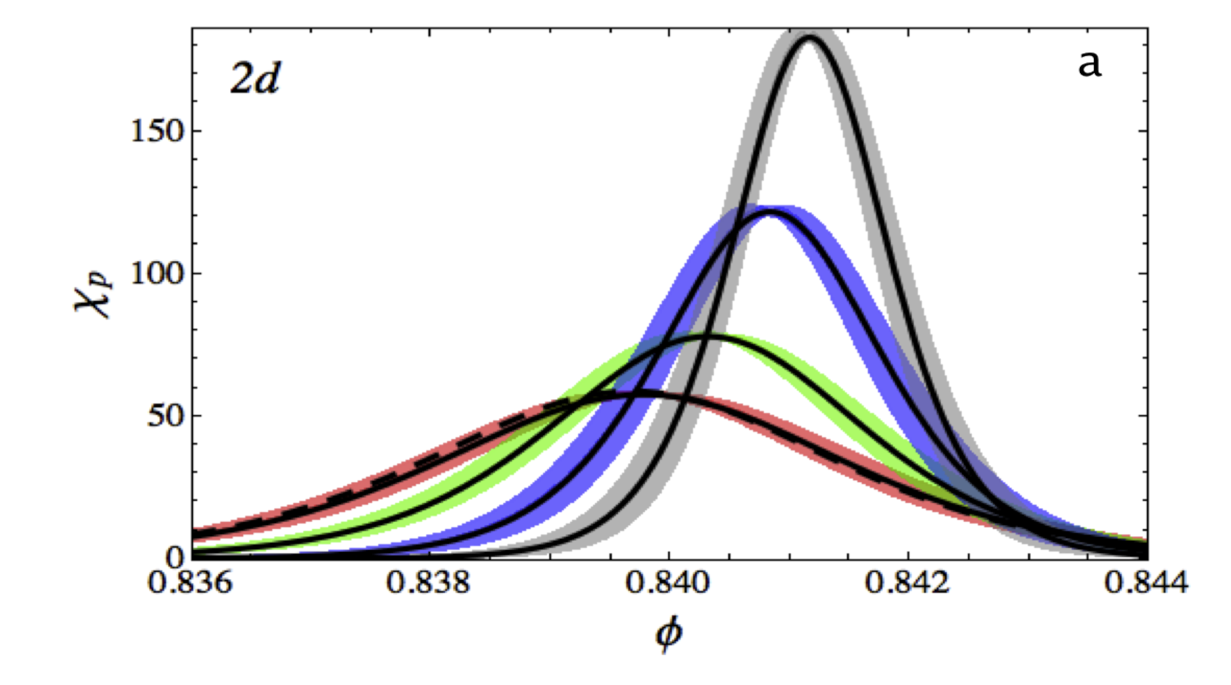}
\includegraphics[width=3.0in]{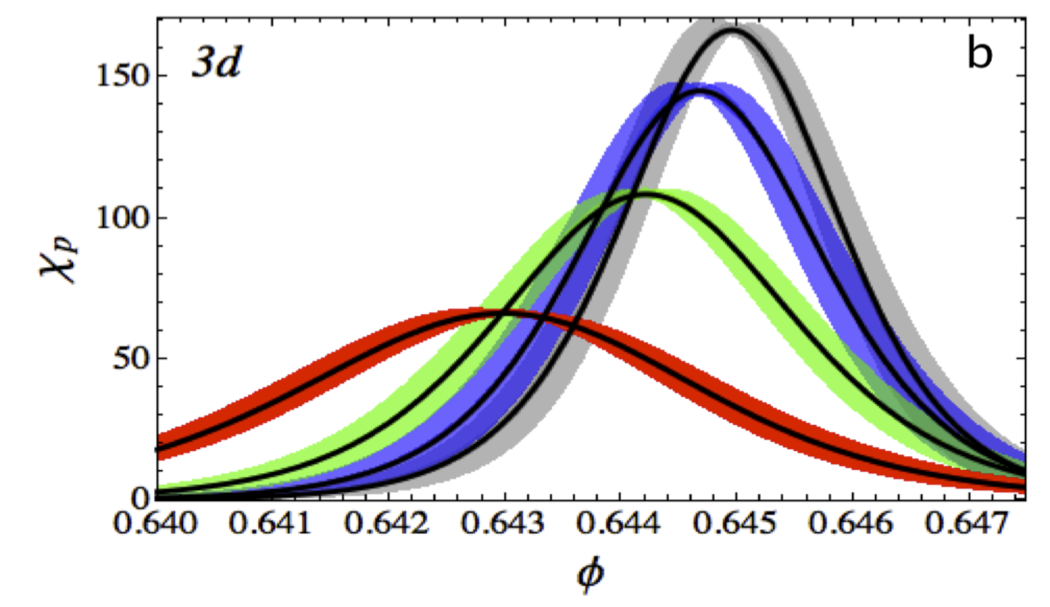}
\includegraphics[width=3.4in]{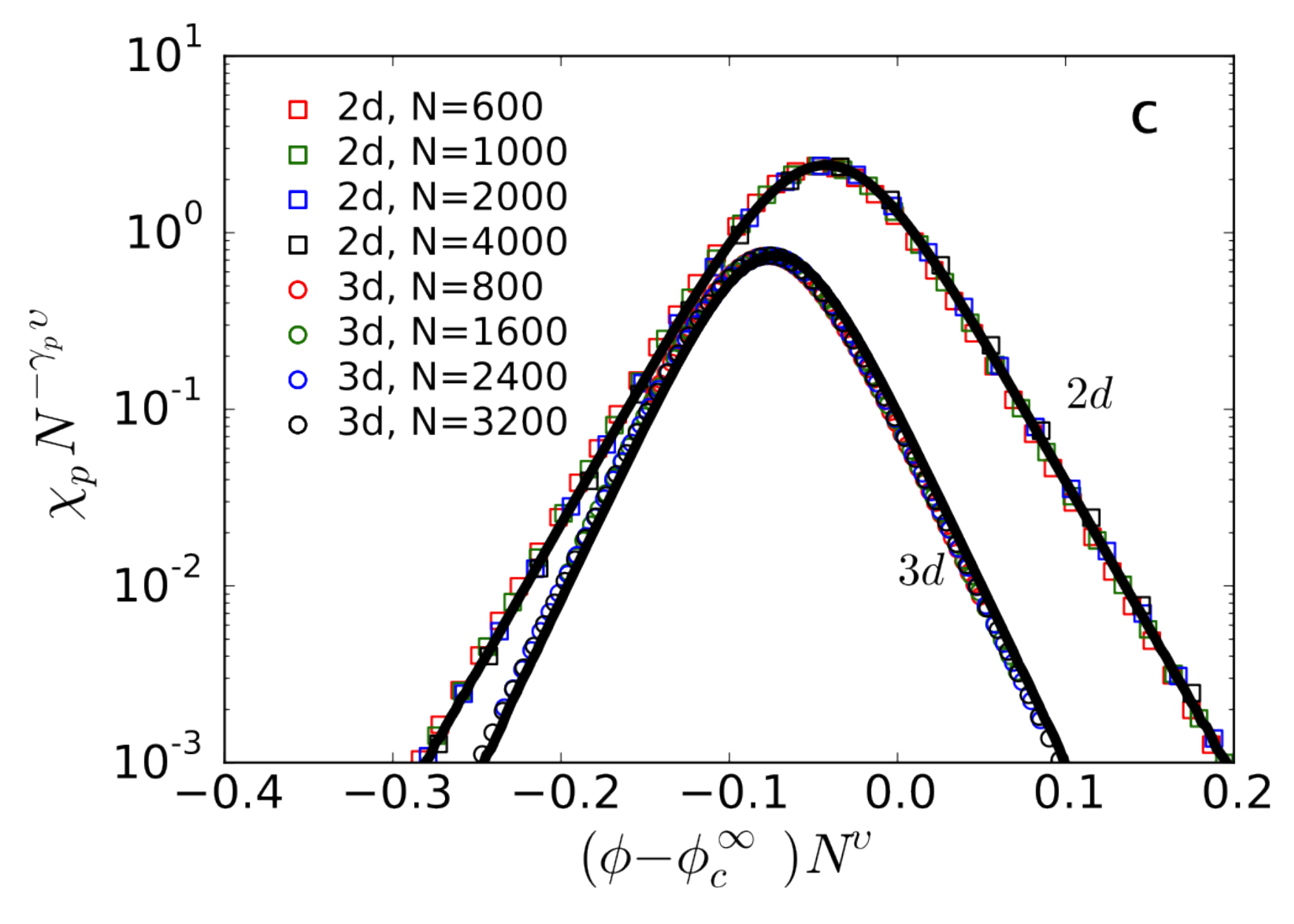}
\label{fig:chi_p}
\caption{ a: Susceptibility calculated as finite difference  for $d=2$.  $N = 600,  1000, 2000, 2000$  with errors as areas around curves in red, green, blue, grey. For $N=600$, solid  line harmonic repulsion and dashed line Hertzian repulsion; all other values of $N$ show harmonic repulsion.   b: Susceptibility calculated as finite difference for $d=3$.   $N = 800, 1600, 2400,  3200$  in red, green, blue, grey.  c: Universal scaling function (curves, Eq. ~\ref{g}) and finite-difference approximation (points) for nonsingular part of pinning susceptibility, $g_d(x)$, for $d=2, 3$.}
\end{figure}

We also calculate $\chi_p$ for each value of $d$ and $N$ using a finite-difference version of Eq.~\ref{chi_p}: 
$\chi_p  \   \approx  N \frac{\ p_j(N_f^\prime) - p_j(N_f)}{N_f^\prime- N_f}$  for limitingly small values of $N_f^\prime , \ N_f$.  
Using $N_f^\prime=2, \ N_f = 1$ in the finite difference yields smooth curves for  $N=600, 1000, 2000$ and $4000$ in $d=2$ in Fig. 3a, and for $N = 800,1600, 2400$ and $3200$ in $d=3$ in Fig. 3b.  We have additionally calculated $\chi_p$ using pairs of $N_f$ values other than $\{2,1\}$ (not shown).  We find that the finite-difference approximation to $\chi_p$ develops a progressively higher peak as $N_f^\prime \rightarrow N_f$.  Uncertainty in $\chi_p(\phi)$ arises from error in the parameter $a$ (error in $b$ contributes much less) which is fit independently for $p_j(\phi, 2)$ and $p_j(\phi,1)$ before the difference is calculated.  This uncertainty is shown as an ``envelope'' about each curve in Figs. 3a,b.  Note that Fig. 3a contains data for both harmonic (solid line) and Hertzian (dashed) potentials for $N=600$.  The disagreement  between the two curves is significantly smaller than the error for either curve, supporting the expectation that the pinning susceptibility near criticality is independent of the details of the repulsive potential. 

Eqs. ~\ref{scaling_function} and ~\ref{g}  arise from differentiating the scaling form for $p_j$  in Eq.~\ref{two_variable} with respect to its second argument.  In Fig. 3c, we show the universal functions $g_d$ as curves for $d=2,3$.  Finite-difference results for different $N$ are shown as points for selected values of $\Delta \phi N^{\upsilon}$. There is excellent agreement between the points and the curves, indicating that the data at each $N$ are in good agreement with the fitted parameter values in Table I obtained by fitting to data at all $N$ and $N_f$. The universal function $g_d$ peaks at $x = -0.043, \ -0.078$ for $d=2, 3$, respectively.  The scaling form of Eq.~\ref{scaling_function} implies that in the thermodynamic limit, we obtain a power-law divergence of $\chi_p \sim | \Delta \phi | ^{-\gamma_p}$.

Note from Table I that the values of $\phi_c^\infty$ are in excellent agreement with previous work on bidisperse soft spheres ~\cite{OHern2002, Vagberg2011}. The finite-size exponent $\upsilon$ for $d=2$ and $d=3$ is consistent within uncertainty, as expected for systems at or above the upper critical dimension.    It is also consistent with the central-limit-theorem value of $\upsilon=1/2$, identified earlier in the absence of pinning in Ref.~\cite{epitome2003}, and with $\upsilon=0.465\pm 0.01$ obtained for $d=2$ systems by V{\aa}gberg, et al.~\cite{Vagberg2011}, who included power-law corrections to scaling in their analysis.  

The pinning susceptibility exponent $\gamma_p$ in $d=2$ and $d=3$ is significantly different when we consider only statistical errors -- in contrast to the dimension-independent values expected above the upper critical dimension of two.  However, these do not include systematic errors due, say, to choice of theoretical analyses.  For example, one way of including logarithmic corrections in $d=2$, the expected upper critical dimension, leads to a considerably higher value of $\gamma_p = 1.50 \pm 0.95$ (statistical errors). This agrees well with $\gamma_p \sim 1.53$ in $d=3$, but one must recognize that the estimated range of systematic errors is enormous.  As a proxy for exploring different theoretical models, Ref. \cite{Mortensen2005} proposed a method which explores the range of fits that is comparable in residual to the best fit.  Following their prescription, we find much larger systematic uncertainties in our estimates of $\gamma_p$ (Table I).  Therefore, our numerical results cannot resolve whether $\gamma_p$ is the same in $d=2,3$.  

Indeed, one can argue that $\gamma_p$ may depend on $d$ as well as a $d$-independent exponent, $\nu$.  
Conceptually, the narrowing of the jamming transition with increasing system size
~\cite{epitome2003} and the shift in the average transition ~\cite{Reichhardt2012, Brito2013}, imply a derivative of the jamming probability which depends singularly on the density of frozen particles.  
For attractive pins in $d=2$~\cite{Reichhardt2012}, it was noted that average distance between pins, $l_f$  could be equated with a correlation length  $\xi \propto \Delta \phi^{-\nu} $
at the jamming threshold.   Since $l_f \propto n_f^{-1/d}$, this argument  suggests $\Delta \phi \propto n_f^{1/d \nu}$ as a scaling relation.  Our Eq.~\ref{two_variable} would thus be written $p_j = F( \Delta \phi \ N^{\upsilon} , n_f  N^{ \upsilon d \nu} ) $, 
implying $\gamma_p \equiv d \nu$.  With $\nu = 1/2$ in $d=2,3$, this $d$-dependent prediction for $\gamma_p$ is consistent with numerical results of Table I. 

In summary, we have found that jamming is infinitely susceptible to pinning at the jamming transition in the thermodynamic limit.   We have identified a new exponent associated with power-law divergence of this pinning susceptibility. The divergent response to pinning, even in the limit of infinitely dilute pinning, suggests that it should be fruitful to study the interplay of jamming and pinning at higher pinning fractions.  

We thank A. A. Middleton for the scaling argument leading to $\gamma_p = d \nu$, and thank R. Kenna for instructive discussions.  Acknowledgement is made to the donors of the American Chemical Society Petroleum Research Fund for support of this research, and to the Division of Natural Sciences and Provost's Office of Swarthmore College 
 (ALG, SN, EP). This research was also supported by the US Department of Energy, Office of Basic Energy Sciences, Division of
Materials Sciences and Engineering under Award DE-FG02-05ER46199 (AJL, CPG), and by
the National Science Foundation under award DMR 1312160 (JPS). This work was partially supported by a Simons Investigator award from the Simons Foundation to AJL and by a University of Pennsylvania SAS Dissertation Award to CPG.  ALG was the recipient of a Eugene M. Lang Faculty Fellowship from Swarthmore College in support of sabbatical leave.




\end{document}